\documentclass[12pt]{iopart}
\usepackage{graphicx}

\newcommand{\BEC}{Bo\-se-Ein\-stein con\-den\-sate}
\newcommand{\text}[1]{{\rm #1}}
\newcommand{\IM}[1]{{I_{\text{M#1}}}}
\newcommand{\IH}[1]{{I_{\text{H#1}}}}

\begin{document}

\title{Transporting, splitting and merging of atomic ensembles in
 a chip trap}

\author{P~Hommelhoff\footnote[1]{now at Department of Physics, Stanford
University, Stanford, California 94305-4060, USA},
W~H{\"a}nsel\footnote[2]{now at Institut f{\"u}r
Experimentalphysik, Universit{\"a}t~Innsbruck, Technikerstr.~25,
A~6020~Innsbruck, Austria}, T~Steinmetz, T~W~H{\"a}nsch and
J~Reichel}

\address{Max-Planck-Institut f{\"u}r Quantenoptik and
Ludwig-Maximilians-Universit{\"a}t M{\"u}nchen,
Schellingstra{\ss}e~4, D~80799~M{\"u}nchen, Germany}

\ead{peter.hommelhoff@physik.uni-muenchen.de}

\begin{abstract}
  We present a toolbox for cold atom manipulation with time-dependent
  magnetic fields generated by an atom chip. Wire layouts, detailed
  experimental procedures and results are presented for the following
  experiments: Use of a magnetic conveyor belt for positioning of cold
  atoms and Bose-Einstein condensates with a resolution of two nanometers;
  splitting of thermal clouds and BECs in adjustable magnetic
  double well potentials; controlled splitting of a cold reservoir.
  The devices that enable these manipulations can be combined with
  each other. We demonstrate this by combining reservoir splitter and
  conveyor belt to obtain a cold atom dispenser.  We discuss the
  importance of these devices for quantum information processing, atom
  interferometry and Josephson junction physics on the chip. For all
  devices, absorption-image video sequences are provided to
  demonstrate their time-dependent behaviour.
\end{abstract}


\submitto{\NJP}

\section{Introduction}
Recent progress in quantum engineering with neutral atoms
crucially depends on the ability to tailor sophisticated
potentials for manipulating the atoms. For instance, the
implementation of a strongly confining three-dimensional optical
lattice potential in a Bose-Einstein condensate (BEC) apparatus
lead to the observation of the Mott--insulator
state~\cite{Greiner02}.  Neutral-atom quantum information
processors and guided-wave interferometers will require even more
complex potentials. Thus, for example, a qubit conveyor belt has
been envisioned \cite{DiVincenzo00} to transport qubit atoms to
the computation area without affecting the qubit state.
Guided-wave interferometers will need continuous atom lasers
\cite{Chikkatur02} as a source, and require highly stable,
coherent beam splitters---currently a subject of active research
in the field of microchip-based atom traps (``atom chips'').
Indeed, atom chips are particularly well suited for this kind of
complex atom manipulation, since highly complex trapping
potentials can be created with lithographically produced
conductors on a chip \cite{Reichel02APB,FolmanReview02}. As in
microelectronics, complex features may be obtained from simpler
building blocks. A few such building blocks have been demonstrated
and described in the last four years, such as on-chip BEC
sources~\cite{Ott01,Haensel01BEC,Schneider03}, conveyor belts for
ultracold atoms~\cite{Haensel01Conv} and BECs~\cite{Haensel01BEC},
guides~\cite{Mueller99,Dekker00}, Y-shaped
connections~\cite{Cassettari00} and a switch for ultracold
atoms~\cite{Mueller01} and also BEC guides \cite{Leanhardt02}.
Here we describe the use of time-dependent magnetic trapping
potentials to achieve the following functions:
\begin{enumerate}
\item {\it Nanopositioning} of atoms with a resolution of two
  nanometers.  Taking our atomic conveyor belt \cite{Haensel01Conv} as a
  starting point, we use numerical optimization of the time-dependent
  currents to achieve transport of a strongly confining trap with
  minimum deviation from a straight line, so that a BEC can be
  transported and accurately positioned with this conveyor belt.  A
  detailed analysis of positioning errors complements the experimental
  results.

\item {\it Splitting} of atom clouds and BECs by reversible
  transformation between single and double-well potentials. We
  demonstrate splitting and merging of thermal clouds, and splitting
  of a BEC.  This first demonstration of on-chip splitting of a
  three-dimensionally trapped BEC provides important experimental
  input for the realization of trapped-atom
  interferometers~\cite{Hinds01,Haensel01Interf, Shin04} 
  and for collisional  phase gates~\cite{Calarco00}.

\item {\it Cold atom dispensing}, i.e. extraction of a small quantity
  of trapped atoms from a larger reservoir. We show how this function
  can be combined with the conveyor belt to transport the dispensed
  atom cloud from the reservoir to a destination elsewhere on the
  chip. Finally, we discuss the use of this device to pump a
  continuous atom laser.

\end{enumerate}

\section{Experimental procedure}

\begin{figure}
\centering{
  \includegraphics[width=6cm]{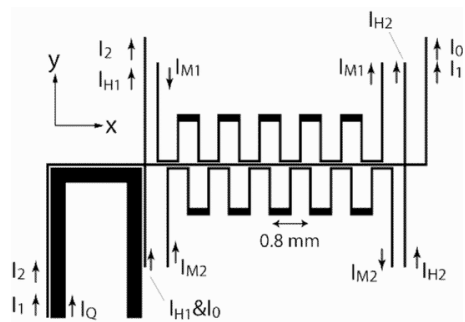}
} \caption{Layout of the gold conductors on the atom chip. A
  quadrupole field for the last stage of the mirror-MOT can be
  produced by a current $I_Q$ in the broad U-shaped wire together with
  an external magnetic field along $y$. With appropriate bias fields
  and currents, the atoms can be magnetically trapped and transferred
  to the right section. There, meandering wires are used to create a
  chain of magnetic potential wells in which the atoms can be
  transported as on a conveyor belt.}
\label{pic:Substrat}
\end{figure}

Our general setup and initial chip trap loading procedure are
detailed in~\cite{Reichel01} and summarized below. All experiments
described here have been performed with the same chip. For each
experiment, the atoms have been loaded from a
mirror-MOT\footnote{MOT: magneto-optical trap}~\cite{Reichel99},
which involves a reflective layer on top of the lithographic
wires. For the various experiments, different loading regions have
been used, as described later in this section. 

The trapping potential is generated by superimposing a homogeneous
external magnetic field with the magnetic field created by the
current carrying wires on the chip. Depending on the wire
geometry, both quadrupole and Ioffe-Pritchard traps can be created
(for an overview see ~\cite{Reichel02APB,FolmanReview02}). The
current carrying wires are lithographically produced gold
conductors on an aluminum nitride ceramic chip
(\fref{pic:Substrat}). This chip is mounted upside down in a glass
cell, which is evacuated to a base pressure in the 
$10^{-10}\,$hPa range. We typically capture $5\cdot10^6$
$^{87}$Rb~atoms close to the chip surface in the mirror-MOT,
perform a 3\,ms-stage of polarization gradient cooling and
optically pump the atoms into the doubly-polarized state $F=2,m_F
= 2$ within 1\,ms. From here, we use two different procedures to
load the atoms into the magnetic trap, depending on whether we
want to cool the atoms evaporatively in the magnetic trap or not.
A detailed description of the two final loading steps will be
given in the next two subsections.

Our chip, depicted in \fref{pic:Substrat}, consists of two main
sections. In the left section, the currents $I_Q$ and $I_2$ can be
used to create a large-volume quadrupole and Ioffe-Pritchard trap,
respectively. In the right section with the meandering currents
$I_\text{M1}$ and $I_\text{M2}$, moving potential wells can be
created to transport the trapped atoms. The two straight wires
H$_1$ and H$_2$ to the left and to the right of the conveyor
section can be used to create stationary magnetic traps and
barriers.

\subsection{Loading into the large-volume trap}
\label{sec:Loading}

\begin{figure}
  \centering{ \includegraphics[height=10cm]{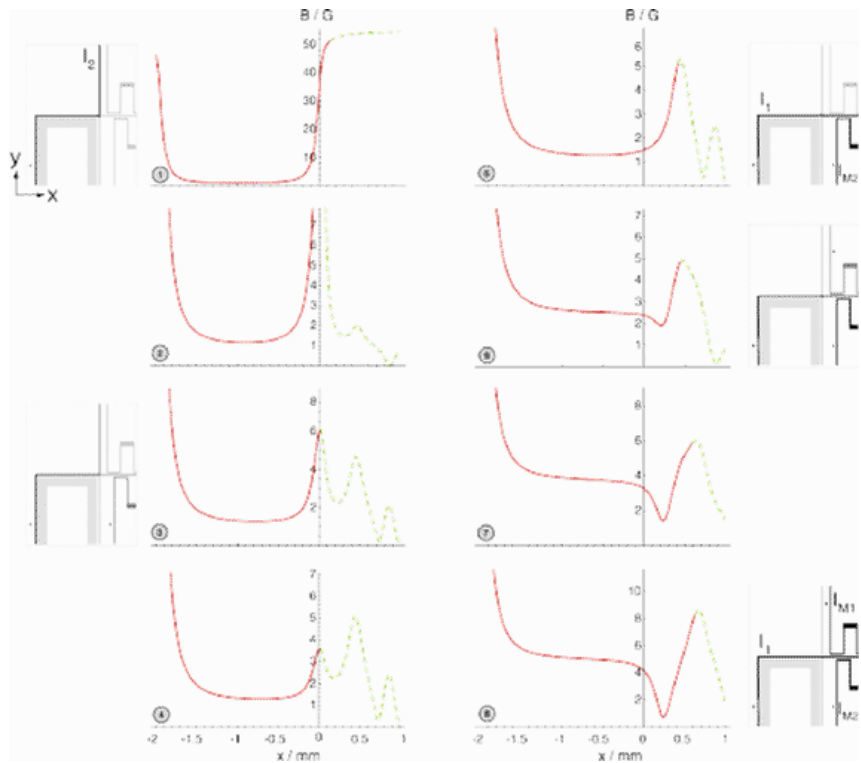} }
  \caption{Plots of the magnetic field minimum along $x$ illustrating
    the transformation of the initial trap into the conveyor start
    trap. The relevant conductors for each phase are highlighted in
    the insets. The first image shows the compressed trap ($I_2 =
    2\,$A, $B_y=55\,$G), in which evaporative cooling is performed (1).
    In order to shift the atoms to the starting position of the
    conveyor belt, located below the wire M1, we perform the following
    steps: $I_1$ is ramped up to 2\,A while $I_2$ is decreased with
    $I_1 + I_2 = 2\,\mathrm{A}= \mathit{const}$. At the same time,
    $I_{\text{M2}}$ is raised to 1\,A ($(2) \dots (5)$ within
    100\,ms). During this process, the trap is expanded to the right.
    We then produce a `dimple trap' by ramping up $B_x$ to 7\,G and
    locally reduce that field by sending 1\,A through M1 ($(6) \dots
    (8)$ within 150\,ms). This produces the desired trap at the
    conveyor starting position. All ramping times are chosen such that
    the atoms can follow the minimum of the trapping potential (full
    line) without significant decrease of phase space density.}
\label{pic:IoffeToMotor}
\end{figure}

For applications which require very cold or even Bose condensed
atoms, a trap with a large initial volume is necessary, since
typically $99\,\%$ of the atoms are lost in the course of
evaporative cooling to BEC. We use a current $I_Q$ in the U-shaped
wire together with the offset field $B_y$ to create and compress a
mirror-MOT of large volume (see \fref{pic:Substrat}). After
polarization gradient cooling and optical pumping, an elongated
Ioffe-Pritchard trap is created with $I_2=2\,$A and a bias field
$B_y = 8\,$G. In this trap, we capture up to $3\cdot 10^6$ atoms
at a typical temperature of 45$\,\mu$K. If lower temperatures are
required, we perform radio-frequency induced evaporative
cooling~\cite{Haensel01BEC}. To achieve a high elastic collision
rate, we compress the trap by ramping $B_y$ to 55\,G within
300\,ms. We then relax the trap by reducing the offset field to
40\,G and, if needed, perform a second stage of evaporative
cooling. \Fref{pic:IoffeToMotor} describes the steps that transfer
the atoms from this trap into the conveyor.

\subsection{Direct loading into the conveyor belt}
\label{sec:DirectLoading}

\begin{figure}
\centering{
\includegraphics[height=4.5cm]{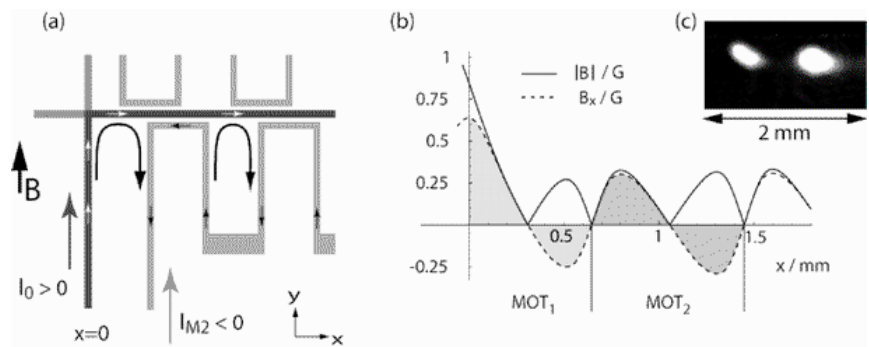}
} \caption{Direct loading into the conveyor belt: in the last
mirror-MOT phase, we produce neighbouring quadrupole fields with
$I_0 = 0.3\,$A, $I_{\text{M2}} = -0.3\,$A and $B_y = 1.5\,$G. In
(a), the arrows indicate the effective U-shaped current flow
responsible for the quadrupole field. In (b), the axial component
of the magnetic field and its modulus are shown, revealing the
successive quadrupole minima along the $x$-axis. We load the atoms
from the MOT into the magnetic conveyor traps (see
\fref{pic:BECMotor} (a)) by switching on currents $I_0 = 2\,$A and
$I_{\text{M1}} = 1\,$A, and fields $B_x = 7\,$G, $B_y = 16\,$G.
(c) Absorption image of two atom clouds that were loaded into the
two leftmost trapping potentials (the left cloud contains
$110\,000$, the right $260\,000$ atoms). The image is taken
$0.15\,$ms after switching off the trapping potential.}
\label{pic:LoadConveyor}
\end{figure}

Alternatively, atoms can be loaded directly into the conveyor
potential. In this case, quadrupole fields for the last MOT stage
are created below the meandering wires as shown in
\fref{pic:LoadConveyor}. After polarization gradient cooling and
optical pumping, the magnetic conveyor is switched on as detailed
in the figure caption. The volume of these traps is comparatively
small, such that loading at typical MOT densities limits the atom
number to a few $10^5$ per well. Temperatures of 10 to 50\,$\mu$K
and densities on the order of $10^{10}\,\text{cm}^{-3}$ are
obtained in this way.

\section{Transport and nanopositioning} \label{sec:conveyor}

The atomic conveyor belt was proposed in~\cite{Reichel99} and
first demonstrated in~\cite{Haensel01Conv}. In its simplest
implementation, sinusoidally modulated currents are used to
displace the `bins' containing the atoms. In this case, the bins
follow an overall straight path with a superimposed periodic
deviation, as described in~\cite{Haensel01Conv}. Here we show how
numerical optimization of the modulated currents and bias fields
can be used to reduce the periodic deviation, such that ideal
linear transport is approximated with micrometric accuracy. This
optimization was crucial to achieve the goal of non-destructive
BEC transport~\cite{Haensel01BEC}.

\subsection{Basic conveyor belt} \label{sec:MotorBasic}

\begin{figure}
\centering{
\includegraphics[width=8cm]{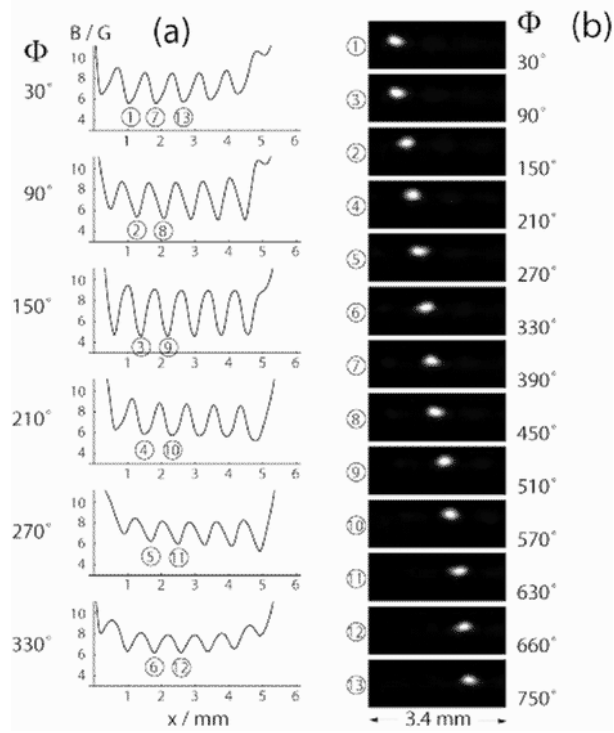}
} \caption{Calculated potential (a) and absorption images of
$150\,000$ atoms in the basic conveyor belt (b). The field modulus
given in (a) is the minimum value in the $y$-$z$~plane.}
\label{pic:MotorPotAndPics}

\end{figure}

The conveyor belt consists of two main components
(\fref{pic:Substrat}). A 2D-Ioffe-Pritchard potential is created
by the current $I_0 = 2$\,A and an external bias field
$B_y=16$\,G. This potential traps the atoms in the $y$-$z$~plane.
The confinement along $x$ is accomplished by the meandering
currents $I_\text{M1}$ and $I_\text{M2}$ and an external bias
field $B_x=7$\,G. The resulting potential consists of a chain of
minima as shown in \fref{pic:MotorPotAndPics}. When the currents
are modulated according to
\begin{equation} (I_{\text{M1}}, I_{\text{M2}}) =
1\text{\,A\,}(\cos{\Phi}, -\sin{\Phi}),
\end{equation} these minima move continuously to the right with increasing phase angle $\Phi.$
In the experiment of \fref{pic:MotorPotAndPics}, $\Phi$ is varied
linearly with time, $\Phi = \omega t$ with $\omega = 2 \pi /
150\,$ms. We call this mode of operation the ``basic conveyor
belt''. Given the spatial period of $800\,\mu$m, the atoms are
transported at a velocity of $5.3\,$mm/s and traverse the full
conveyor length of $4.6\,$mm within 863\,ms. At this velocity, we
do not observe losses due to the transport, and the heating is
below our temperature resolution of $2\,\mu$K.

\begin{figure}
  \centering{ \includegraphics[height=3.5cm]{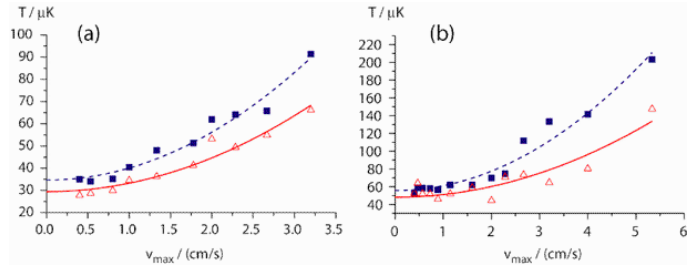}}
  \caption{The temperature of the atom cloud after a transport distance of
    $3.2\,$mm is plotted versus the maximum transport velocity. In
    (a), the basic conveyor belt version described in
    \sref{sec:MotorBasic} has been used, in b), the height-optimized
    version of \sref{sec:HeightOptMot}. The cloud was accelerated and
    decelerated over the first and last $0.8\,$mm, respectively. The
    squares indicate the temperature in transport direction
    ($x$-direction), while the triangles denote the temperature along
    $z$-direction.  The difference along the axes shows that the atoms
    have not yet thermalized after the heating process. The lines are
    square fits to the data, yielding a temperature increase of
    $\Delta T_x = 5.4\, \mu \text{K} \cdot (v_{max} /
    (\text{cm}/\text{s}))^2$ and $\Delta T_z = 3.8\, \mu \text{K}
    \cdot (v_{max} / (\text{cm}/\text{s}))^2$ for the basic conveyor.
    For the height-optimized conveyor, slightly smaller heating of
    $\Delta T_x = 5.5\, \mu \text{K} \cdot (v_{max} /
    (\text{cm}/\text{s}))^2$ and $\Delta T_z = 3.0\, \mu \text{K}
    \cdot (v_{max} / (\text{cm}/\text{s}))^2$ has been observed
    (fitting errors of all $\Delta T$ are $0.4\,\mu \text{K} \cdot
    (v_{max} / (\text{cm}/\text{s}))^2$).} \label{pic:MotorHeating}
\end{figure}

However, if we increase the transport velocity to more than
$1\,\mathrm{cm}/\mathrm{s}$, the atom cloud is measurably heated
(\fref{pic:MotorHeating}). Two effects contribute to this heating:
First, deviations of the trap position from the ideal straight
line cause small periodic accelerations. Second, shape and
steepness of the trap also vary periodically. Both heating effects
can be reduced, or even completely avoided, by optimizing the time
dependence of the various currents and bias fields. This is
demonstrated in the next two subsections.

\subsection{Height optimized version} \label{sec:HeightOptMot}

\begin{figure}
  \centering{
    \includegraphics[height=7cm]{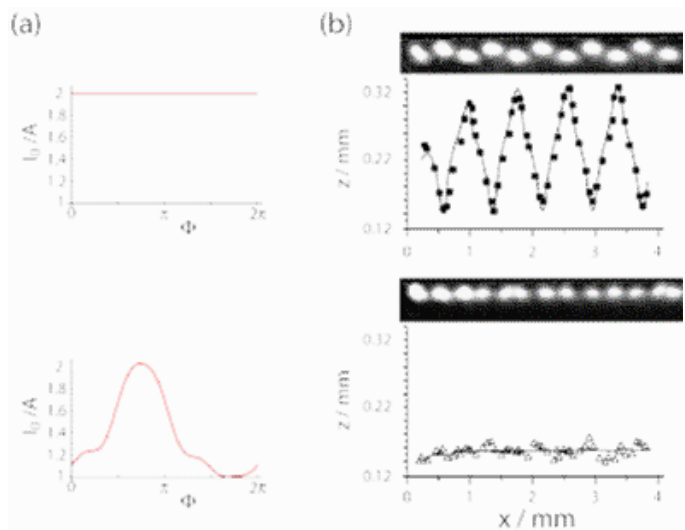}
  }
  \caption{Comparison of the basic conveyor (top) and the
    height-optimized version (bottom). In (a), the current $I_0(\Phi)$
    is shown for the two cases. Part (b) illustrates the height
    variation during transport, $z(x)$. Squares and triangles are
    experimental results (center-of-mass positions extracted from the
    absorption images). The lines represent the theoretical prediction
    from magnetic field simulations, with a constant offset $z_0$
    added to account for experimental uncertainty in the surface
    position.} \label{pic:MotorHeightAll}
\end{figure}

In the basic version of the conveyor discussed above, one obvious
deviation from the ideal linear transport is the variation in
height. The experiments and simulations in the upper row of
\fref{pic:MotorHeightAll} show that the excursions are of the
order of $\pm75\,\mu$m. This effect can be understood from
\fref{pic:Substrat}: once per period, the potential minima pass
through a position where both meandering wires are close to the
central wire. This happens for $\Phi=3\pi/4$, when both currents
$I_\text{M1}$ and $I_\text{M2}$ are running antiparallel to the
main current $I_0$. The atoms then feel a smaller effective
current $I_{\text{eff}}\sim I_0+\IM1+\IM2$ along $x$. Since the
trap-surface distance is proportional to $I_{\text{eff}} / B,$ the
atoms move closer to the wire at these positions. This vertical
excursion can be avoided if the central current $I_0$ and/or the
field $B_y$ are adjusted appropriately.

For our experiments, we have chosen to keep $B_y$ constant and to
vary $I_0$. A numerical optimization of $I_0(\Phi)$ has been used
to keep the height of one particular well constant during one
shifting period. The resulting current is plotted in
\fref{pic:MotorHeightAll}~(a).
The experimental results \fref{pic:MotorHeightAll} (b) show that
the height is indeed constant to $\pm 15\,\mu$m, even over the
whole conveyor length.

Despite this optimization of the height, the heating rate at high
transport velocities is only slightly reduced
(\fref{pic:MotorHeating}). This is due to the fact that the
deviation in the $y$-direction has been left uncompensated.
Furthermore, as mentioned above, the shape of the trapping potential
changes significantly during the transport. This also appears in the
variation of the axial and transverse trapping frequencies,
$90\,\text{Hz} < \nu_{\parallel} < 230\,\text{Hz}$ and
$440\,\text{Hz} < \nu_{\perp} < 670\,\text{Hz}$, which characterize
the potential at the bottom of the trap. For very cold ensembles of
atoms, the conveyor can be further optimized, as described in the
next subsection.

\subsection{Optimized BEC transport}

\begin{figure}
\centering{\includegraphics[height=4cm]{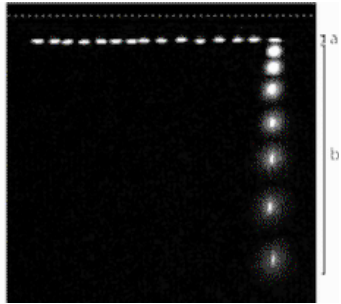}}
\caption{Transport of a \BEC\ in the fully optimized conveyor
  belt. (a) The condensate is produced on the left and transported to
  the right; (b) time-of-flight images after the transport prove that
  the condensate survived: the characteristic bimodal distribution is
  clearly visible. The dashed line indicates the position of the chip
  surface.}
\label{pic:BECMotor}
\end{figure}

When the atom cloud is so cold that it remains within the parabolic
region of the potential, the task to keep the potential shape constant
reduces to keeping the three trap frequencies constant. The two
transverse frequencies are very similar and are mainly given by the
2D-Ioffe-Pritchard potential. Therefore, we can influence them by
changing the current $I_0$ and the offset fields $B_x$ and $B_y$.  The
axial frequency is due to the meandering currents $I_\text{M1}$ and
$I_\text{M2}$ and can be adjusted by changing the amplitude of the
currents. In a final step, we take into account that the trap position
along $x$ is not an exactly linear function of $\Phi$.

For every $x$, we have numerically calculated appropriate values of
$B_x, B_y, \IM1, \IM2$ to obtain a trap at position $x$, $z = (260 \pm
1)\,\mu$m with frequencies $\nu_{x} = (60 \pm 1.5)\,$Hz and $\nu_z =
(400 \pm 2.5)\,$Hz. $\nu_y$ is coupled to $\nu_z$ and varies between
378 and 393\,Hz over the full range of $x$ values. $y$ was not taken
into account in the optimization because we cannot modulate $B_z$ in
our current setup.

With this optimization, we were able to transport a \BEC\ over a
distance of $1.6\,$mm within 100\,ms, i.e.  at $\bar{v} =
1.6\,\text{cm}/$s (\fref{pic:BECMotor}). To avoid sloshing of the
condensate, we accelerated and decelerated the conveyor using a
Blackman pulse in the velocity~\cite{Blackman58}:
\begin{equation} v(t,
  T) = \frac{1}{T} \left[ 1 - \frac{25}{21} \cos\left(2 \pi
      \frac{t}{T}\right) + \frac{4}{21} \cos\left(4\pi
      \frac{t}{T}\right)\right]
\end{equation} with $T = 50\,$ms.


\subsection{Positioning accuracy}

An important application for the atomic conveyor belt is precise
positioning of an atom (cloud) inside an optical resonator.  For
cavity-QED experiments, the atom-field coupling crucially depends
on the overlap between the atomic wavefunction and the cavity
mode.  For these experiments as well as for scanning atomic
microscopy~\cite{Lin04}, positioning accuracy of the trap center
is essential. It is interesting to evaluate the position accuracy
for a chip trap under realistic experimental conditions, i.e. in
the presence of fluctuating fields, field gradients and currents.
As a realistic example, we consider a trap with the highest
transverse trapping frequency that we have achieved in our setup,
which is 11\,kHz. (Much higher frequencies are possible with
straightforward technical improvements -- see
e.g.~\cite{Reichel02APB}.)  The parameters for this trap are $I_1
= 2\,$A, $I_{\text{M1}} = 1\,$A and ${\bf B} = (5, 60, 0)\,$G; the
trap is located $73\,\mu$m from the chip surface. The transverse
ground state extension ($1/e$ radius) in this trap is 100\,nm. The
axial trapping frequency and the axial ground state extension are
730\,Hz and 400\,nm, respectively.

If we assume a relative current stability of $10^{-5}$ in all
current sources that are used to create the trapping fields, a
simulation yields a position jitter of
$\Delta z_{\textrm{rms}}=1.2\,$nm. Furthermore, the trap position
is sensitive to fluctuating ambient magnetic fields in the
$yz$-plane and to the component $\partial B_x / \partial x$ of an
ambient magnetic field gradient. Their contribution to the
displacement amounts to $\Delta r_{yz}=\delta B_{yz}\times
1.5\,\textrm{nm}/(\textrm{G}/\textrm{cm})$ and $\Delta x=(\partial
B_x / \partial x) \times
0.22\,\textrm{nm}/(\textrm{mG}/\textrm{cm})$. Ambient field
stabilities of $1\,\textrm{mG}$ and $1\,\textrm{mG}/\textrm{cm}$
can be reached with relatively small effort (in atomic fountain
clocks, residual field fluctuations and gradients are several
orders of magnitude smaller than this \cite{Bize99}).  Therefore,
an overall positioning accuracy of a few nm is certainly
realistic. This is almost two orders of magnitude smaller than the
trap's ground state extension and more than two orders of
magnitude smaller than the main optical transition wavelengths for
Rb ($\lambda=780\,$nm and 795\,nm).  Thus, the assumed stability
certainly suffices for applications which require a position
accuracy in the $\lambda / 100$ range. Note that the position
jitter from current fluctuations only affects the $z$ coordinate.
For external field fluctuations below 0.2\,\textrm{mG}, this is
the main contribution to the overall position jitter -- the
position in the $xy$-plane is as stable as $\Delta
x_{\textrm{rms}} = 0.2$\,nm and $\Delta y_{\textrm{rms}} =
0.3$\,nm.

\subsection{Outlook}
With this optimization, the conveyor belt can now be used for
ultraprecise positioning of BECs and single atoms. This is closely
related to the nanopositioning of single trapped ions that was used
to map out the light field distribution in an optical cavity
\cite{Guthoerlein01}. Indeed, we are currently using a two-layer
variant of the conveyor belt \cite{Haensel01Conv} to position atoms
in the evanescent optical field of a microsphere resonator for
single-atom detection~\cite{Long03}.

\section{Double wells: Divide and unite}
A double well potential with adjustable barrier height is the
central building block for the proposed chip-based collisional
phase gate \cite{Calarco00} and trapped-atom interferometer
\cite{Haensel01Interf, Shin04}, 
but also for a magnetic technique \cite{Haensel01Conv} to
continuously pump an atom laser by replenishing its atom reservoir
\cite{Chikkatur02}, and for Josephson-type effects with BECs
\cite{Giovanazzi00,Williams01,Menotti01,Pu02}.

Below we demonstrate two different current configurations that generate
adjustable magnetic double wells. We use the first configuration to
split and reunite a thermal cloud. By measuring the temperature after
up to five iterations of this process, we show that it is adiabatic to
very good approximation. With the second configuration, we split a
BEC as was first done with optical fields in \cite{Andrews97}. Although
the well spacing in our implementation is still too large to observe
interference fringes, this first on-chip BEC splitting provides
important experimental input for future atom chip implementations of
the phase gate and trapped-atom interferometer mentioned above.

\subsection{Thermal cloud} \label{sec:ThermalSplit}

\begin{figure}
\centering{
\includegraphics[width=9cm]{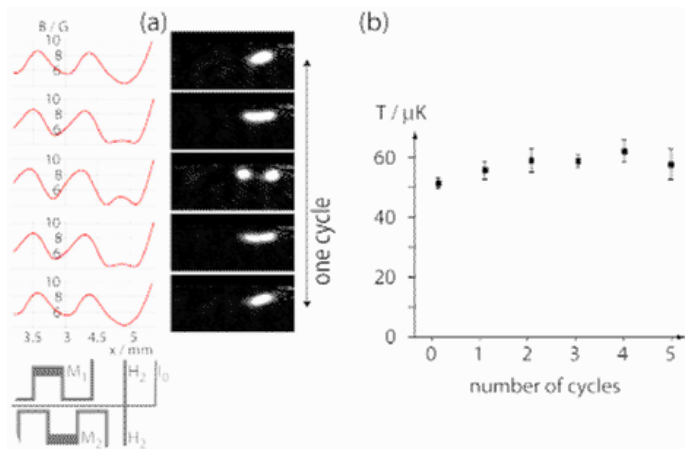}
} \caption{(a) With the help of wire H2, a stationary minimum is
produced  to the right  of the conveyor. Incoming minima from the
left merge with  this  stationary minimum. Reversing the process
leads to a splitting of the trap. We have  split and recombined a
thermal cloud of atoms for a different number of times and
measured the temperature of the combined clouds after  the process
(b). The only feeble increase of the temperature indicates that
the splitting and merging is close to adiabatic.}
\label{pic:SplitThermal}
\end{figure}

We first demonstrate the splitting and merging process with a thermal
cloud positioned at the right end of the conveyor belt
(\fref{pic:SplitThermal}). With the help of a time-dependent current
in the additional wire H2, a stationary minimum is produced below that
wire. Incoming minima from the left are merged with the stationary one
if the current in H2 is driven according to $I_{\text{H2}}(\Phi) =
(0.462 + 0.255 \sin (\Phi + 0.493) - 0.088 \sin (2\Phi - 1.482))\,$A.
All other parameters are controlled as described
in~\sref{sec:MotorBasic}.  The process was designed such that the left
and the right wells have equal phase space volumes during the merging
process.  Reversing this process leads to the splitting of the trap.
In \fref{pic:SplitThermal}~(b) the temperature of the merged cloud is
plotted against the number of splitting and merging cycles.  The
temperature is essentially constant, indicating the adiabaticity of
the process. The slight temperature increase presumably originates
from a slight asymmetry in the splitting ($55 / 45$). We attribute
this asymmetry to an external field gradient along $x$.

\subsection{On-chip splitting of a BEC}

\begin{figure}
  \centering{\includegraphics[height=8cm]{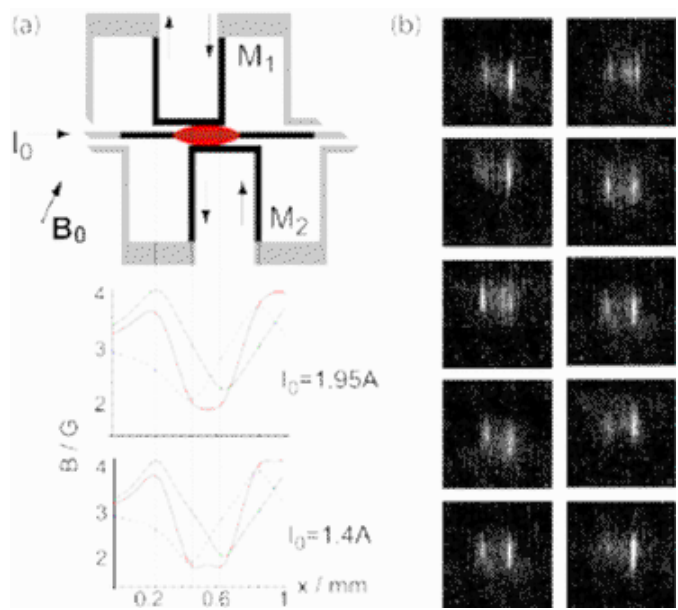}}
  \caption{(a) We split the BEC by lowering the current $I_0$ in the
    central wire. Thereby, the trap center moves from 126 to
    $87\,\mu$m to the surface and a potential barrier, effected by the
    modulation wires, raises from the bottom of the trap. (b)
    Consecutive absorption images of a \BEC\ with 21\,ms
    time-of-flight after it has been split. Clearly, one can see two
    distinct condensates each containing between 500 and 1500 atoms.
    Further details in the text.} \label{pic:PotSplitBEC}
\end{figure}

Fields on the conveyor chip can also be used to split a \BEC.  The
atoms are transported to the position on the belt defined by $\Phi
= 144^{\circ}$ (i.e., close to position 3 in
\fref{pic:MotorPotAndPics}). At this position, the single trap can
be transformed into two traps by decreasing the trap--surface
distance (\fref{pic:PotSplitBEC}). Simulations predict a symmetric
splitting for $\IM1=-283\,$mA and $\IM2=-282\,$mA. In the
experiment we found that the trap splits symmetrically with
$I_{\text{M1}} = -310\,$mA and $I_{\text{M2}} = -274\,$mA. Again,
this deviation presumably results from an external magnetic field
gradient. With a constant external field ${\bf B} = (3, 25, 0)\,$G
we split the initial single trap in two by decreasing the current
$I_0$ from $1.95\,$A to $1.4\,$A within 50\,ms.  After splitting,
the distance between trap centers is $135\,\mu$m and the barrier
height is $60\,\text{mG}\cdot \mu_{\text{Bohr}}$, corresponding to
$4\,\mu \text{K}\cdot k_B$ in our case. Each trap has a mean
frequency $\nu_{\text{HO}} = 680\,$Hz. For 1000 $^{87}$Rb atoms in
each trap, the chemical potential is $130\,\text{nK}\cdot k_B.$ In
this configuration, tunnelling does not occur on relevant
timescales. \Fref{pic:PotSplitBEC}~(b) shows 10 realizations of
this experiment with identical parameters. These results show that
symmetric splitting can be achieved on the chip; however, improved
control over currents and external fields will be needed to obtain
a reproducible splitting ratio. Also, in some images fringes are
visible. These may originate from excitations or from phase
fluctuations that occur just before splitting in the very
elongated trap (cf.~\cite{Dettmer01}). Using a more symmetric wire
pattern \cite{Reichel01} and magnetic shielding will lead to the
improved stability that is a prerequisite for integrated atom
interferometry and quantum information applications.

\section{Cold atom dispenser}

\begin{figure}
  \centering{\includegraphics[height=6cm]{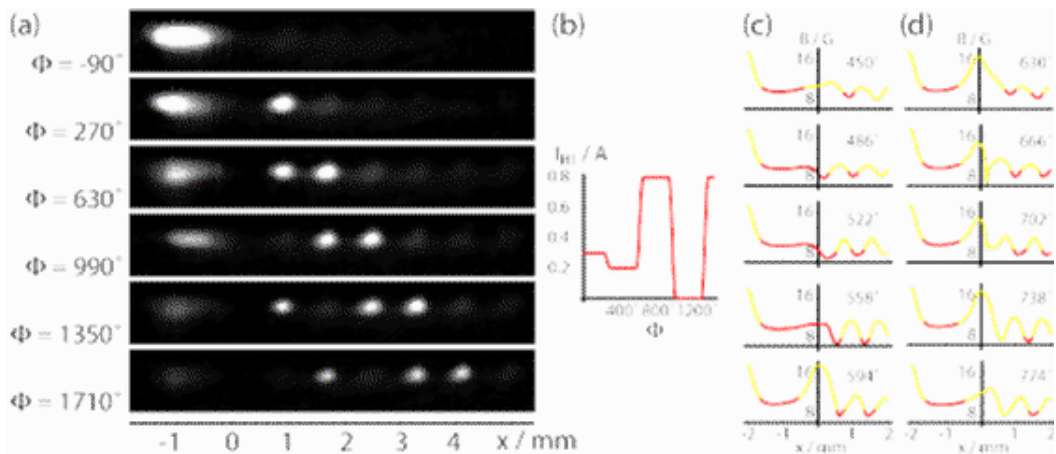}}
  \caption{Cold atom dispenser. (a) Successive absorption images
    illustrate how atoms are extracted from a reservoir of $\sim 10^6$
    atoms. The angle $\Phi$ is related to time by
    $\Phi=2\pi*t/150\,\mbox{ms}$. The current $\IH1$
    (cf.~fig.~\ref{pic:Substrat}) determines the height of the barrier
    between the reservoir and the conveyor and is modulated as shown
    in (b). (c) and (d) show the evolution of the potential in periods
    with and without atom extraction, respectively.}
  \label{pic:Morsen}
\end{figure}

\label{sec:Dispenser} It is a specific advantage of the chip
technology that complex devices can be obtained by combining
simpler building blocks. Here we demonstrate a ``cold atom
dispenser'' that combines the splitting, transport and merging
functions described above. This device may enable the construction
of a continuous atom laser~\cite{Chikkatur02} by purely
micromagnetic means.

The dispenser is fed from a reservoir of trapped atoms to the left
of the conveyor belt. This reservoir is produced by $I_1 =
1.95\,$A, $I_Q = -1.2\,$A, $B_y = 16\,$G and $\IH1=0.3\,$A
(\fref{pic:Substrat}); the role of $H1$ is to create an adjustable
barrier between the reservoir and the conveyor. Switching on the
conveyor belt with $(I_{\text{M1}}, I_{\text{M2}}) =
1\text{\,A\,}(\sin{\Phi}, \cos{\Phi}),$ ${\bf B} = (9,16,0)\,$G
does not extract any atoms from the reservoir yet; but if the
current in H1 is decreased, atoms from the reservoir flood into
departing conveyor belt bins. The fraction of atoms extracted from
the reservoir depends on the height of the barrier.
\Fref{pic:Morsen} shows a reservoir containing $10^6$ atoms.  We
load the first two conveyor bins with $2 \cdot 10^5$ atoms each by
holding $I_{H1}$ at 300\,mA during the first and ramping it down
to 200\,mA during the second splitting (\fref{pic:Morsen} (d)).
(Since the atom number in the reservoir is smaller during the
second splitting, $I_{H1}$ has to be smaller in order to equal
numbers of atoms into both bins). The third conveyor bin is left
empty, which is accomplished by ramping $I_{H1}$ to 800\,mA, and
the fourth conveyor bin is loaded again by ramping $I_{H1}$ to 0.

In a third step, we were able to merge the partial clouds in a
target trap at the end of the conveyor belt. This is shown in the
video sequence {\it Replenish} (see sec.~\ref{sec:video}).  A
continuous atom laser based on this scheme would add an outcoupler
to the target trap. The size of the reservoir needs to be improved
for this application. Surface-induced evaporative cooling
\cite{Reichel99,Harber03} can be used to create a condensate in
the reservoir, or even ``on the fly'' during transport on the
conveyor. In this context, an interesting feature of the conveyor
is that its transport direction can be curved \cite{Long03}. In
conjunction with pinholes, this can be used to protect the
outcoupling region against laser light from the MOT in the loading
region, and to establish a differential vacuum. The atom laser
source would be pumped by the incoming condensates as demonstrated
in~\cite{Chikkatur02}.

\section{Video sequences}
\label{sec:video}

To further illustrate the operation of the different devices we
have created animated versions from series of absorption images.
As our detection scheme is destructive, the trap is reloaded for
each image and the temporal evolution of the parameters is rerun
from the beginning to the time at which the atoms are to be
imaged. Thus, the acquisition of a single image takes roughly
10\,s, and a complete animation is recorded in 30 minutes or less,
depending on the number of pictures.  Computer control of the
experiment and good stability of the setup enable automated
acquisition of such movies.

The first three video sequences {\it ConveyorBasic, ConveyorHoriz}
and {\it BECConveyor} show the different versions of the conveyor
that we discuss in \sref{sec:conveyor}.

The sequence {\it Reunify} illustrates the measurement in which
the adiabaticity of the splitting process was demonstrated (cf.
\sref{sec:ThermalSplit}). Splitting into more than two components
is also possible and is shown in {\it MultiSplit}.

Two modes of operation of the atom dispenser (cf.
\sref{sec:Dispenser}) are shown in the movies {\it Replenish} and
{\it SelectDeliv}. In both versions, the consecutive separation of
atom clouds has been complemented by replenishing the target
reservoir at the end, as would be needed for the implementation of
a continuous atom laser.

Additionally, we provide two further video sequences, which show
the operation of two other on-chip devices that are closely
related to those described above.  For the movie {\it Collider},
two atom clouds were sent to the opposite ends of the conveyor. By
suddenly switching off the currents in M1 and M2, the clouds are
accelerated towards the trap center, where they penetrate each
other.  Since their initial density is ${\cal
O}(10^{10})\,\text{cm}^{-3}$ only, less than $1\,\%$ of the atoms
undergo a collision. With higher initial densities, this behaviour
would change dramatically.  This device is described in more
detail in~\cite{Reichel01}.

Finally, {\it Sixty} gives an example of a more complex atom
trajectory in the plane perpendicular to the substrate. This kind
of transport along a pre-defined trajectory is achieved by
modulating wire currents and external fields according to
functions that were obtained by numerical optimization. More
details on this technique can be found in \cite{Reichel02Sixty}.
The example shown here was recorded on the occasion of the 60th
birthday of one of the authors (TWH).

\section{Conclusion}

We have demonstrated atom chip devices that use time-dependent
magnetic fields for atomic nanopositioning, splitting of thermal
atomic clouds and BECs, and controlled atom dispensing. The potential use of
these devices for atom chip-based quantum information processing, for atom
interferometry with trapped atoms, for studies of Josephson effects
and for pumping of atom lasers has been pointed out.

In order to realize the full potential of future atom chips, some
important problems remain to be solved.  In all the applications
described here, trap-surface spacing was larger than $40\,\mu$m,
so that surface-induced heating and losses \cite{Jones03} did not
play a significant role. If, however, feature sizes approaching
$1\,\mu$m are needed---as is the case for reasonably fast tunnel
coupling between wells---the trap-surface distance will most
likely also approach $1\,\mu$m, and surface effects can no longer
be neglected \cite{Lin04}. High-resistivity materials
\cite{Henkel99b,Harber03} and thinner conductor layers
\cite{Jones03} may be a solution in many cases, especially as
currents scale down with trap-surface distance. Corrugation of
elongated potentials \cite{Fortagh02,Leanhardt02} is currently an
impediment for the realization of proposed guided-wave atom
interferometers \cite{Andersson02}, but is likely to be remedied
by improved manufacturing techniques \cite{Wang04,Esteve04}.
Devices with relatively strong confinement in all three
dimensions, such as the trapped-atom interferometer
\cite{Haensel01Interf}, are generally much less affected by this
problem.

Future atom chips will combine more and more functions on the same
chip to achieve complex tasks. Thus, the next step towards a
chip-based two-qubit gate would be to combine the magnetic double
well with an electrostatic field \cite{Calarco00,Krueger03} or a
microwave field \cite{Treutlein04} to obtain a state-dependent
potential. Combination with optical fields generated by
microoptical elements has also been proposed \cite{Birkl01} and
will further enhance the manipulation capabilities, in particular
when trapping of all magnetic sublevels is required. Combining a
conveyor belt with an optical resonator \cite{Long03} offers
exciting perspectives for single-atom detection and cavity QED
with trapped atoms.

\ack We gratefully acknowledge financial assistance from the
European Union under contracts no. IST-1999-11055 (ACQUIRE) and
IST-2001-38863 (ACQP).

\end{document}